\documentclass[emulateapj,twocolumn]{aastex63}

\usepackage{amsmath}
\usepackage{natbib}
\usepackage{mathtools}
\usepackage{float}
\usepackage{bm}
\restylefloat{table}

\newcommand{\be}{\begin{equation}} 
\newcommand{\ee}{\end{equation}}

\newcommand{\ba}{\begin{eqnarray}} 
\newcommand{\ea}{\end{eqnarray}}

\newcommand{\A}{\mathcal{A}}

\newcommand{\tsec}{t_{\rm sec}}
\newcommand{\aper}{a_{\rm p}}
\newcommand{\eper}{e_{\rm p}}
\newcommand{\apereff}{a_{\rm p,eff}}

\newcommand{\lvec}{{\bm {\hat l}}}
\newcommand{\lpvec}{{\bm {\hat l}}_{\rm p}}

\newcommand{\evec}{{\bm e}}

\newcommand{\ehatvec}{{\bm {\hat e}}}

\newcommand{\evecfull}{{\bm {e}}}
\newcommand{\Lvecfull}{{\bm {L}}}
\newcommand{\Svecfull}{{\bm {S}}}
\newcommand{\p}{{\rm p}}
\newcommand{\D}{{\rm d}}

\newcommand{\svec}{\bm {\hat s}}

\newcommand{\xvec}{\bm {\hat x}}
\newcommand{\yvec}{\bm {\hat y}}
\newcommand{\zvec}{\bm {\hat z}}

\newcommand{\msun}{M_\odot}
\newcommand{\rsun}{R_\odot}

\shorttitle{Tertiary Companion in DI Herculis}
\shortauthors{K.R.\ Anderson \& J.N.\ Winn}

\begin{document}

\title{Were the Obliquities in DI Herculis Excited by an Unseen Tertiary Companion?}

\correspondingauthor{Kassandra Anderson}
\email{kassandra@princeton.edu}

\author[0000-0002-7388-0163]{Kassandra R.\ Anderson} 
\altaffiliation{Lyman Spitzer, Jr.\ Postdoctoral Fellow}
\affiliation{Department of Astrophysical Sciences, Princeton University, Princeton, NJ 08544, USA}

\author[0000-0002-4265-047X]{Joshua N.\ Winn}
\affiliation{Department of Astrophysical Sciences, Princeton University, Princeton, NJ 08544, USA}

\begin{abstract}
The eclipsing binary DI Herculis garnered interest for several decades because of an apparent disagreement between the observed and calculated values of the apsidal precession rate. The problem was resolved when both stars were found to have high obliquities, but the reason for the high obliquities is unknown.
Here we investigate the possibility that the obliquities are (or were)
excited by an unseen tertiary star.
Obliquity excitation in the current orbital configuration can be ruled out with existing
data; any tertiary star
that is sufficiently close or massive to overcome the strong spin-orbit
coupling of the binary would have been detected through
various dynamical effects.
It remains possible that the orbit of DI Herculis was initially wider and
the obliquity was excited during high-eccentricity tidal migration driven by a tertiary companion, but in this scenario it would be difficult to
explain why the observed spin rates are much faster than the pseudo-synchronous rate.
In addition, inward migration is most likely to arise when the mass of the perturbing star is comparable to the binary mass, and such a bright tertiary
would have been detected in 
imaging or spectroscopic data. Alternative explanations that do not invoke a tertiary star should be sought for the large obliquities in DI Herculis.
\end{abstract}


\section{Introduction} 

Measurements of stellar obliquities (spin-orbit angles) may shed light onto the physical
processes governing the formation and evolution of exoplanet systems and binaries
of all kinds. On the exoplanetary front, the discovery of hot Jupiters on severely misaligned orbits has inspired numerous theoretical explanations \citep{winn_fabrycky2015}. Meanwhile, the obliquity information that can be extracted from gravitational-wave signals may help to shed insight into the origin of close black hole binaries \citep[e.g.][]{rodriguez2016}.  Similarly, obliquity information in short-period stellar binaries is relevant to theories for close binary formation. However, at present, obliquity constraints have been obtained only for a handful of short-period stellar binaries.

The DI Herculis system \citep{hoffmeister1930} remains the most extreme known example of spin-orbit misalignment in an eclipsing binary (see the introduction of \citealt{justesen2020} for a recent review of the situation). \cite{albrecht2009} found the primary and secondary stars to have sky-projected obliquities of $\lambda_1 = 72^\circ$ and $\lambda_2 = -84^\circ$
by observing the \cite{rossiter1924}--\cite{mclaughlin1924} effect. Further analysis of the 3D obliquity was performed by \cite{philippov2013}. These obliquity measurements resolved a puzzle that had lasted for several decades over the disagreement between the observed and calculated apsidal precession rates. The calculated apsidal precession rate (with contributions from general relativity (GR), tidal distortion, and rotational oblateness) appeared to be several times faster than the observed value. Various resolutions were proposed \citep[e.g.][]{martynov1980,guinan1985,khaliullin1991,claret1998}, including a failure of general relativity as a last resort. \cite{shakura1985} realized that the calculated apsidal precession rate could be reduced by invoking large spin-orbit misalignments, which would change the direction of the contribution from rotational oblateness. Subsequent analysis constrained the range of obliquities that would be necessary to bring the calculated and observed apsidal precession rates into agreement \citep{company1988,reisenberger1989}. After the obliquity measurements by \cite{albrecht2009}, there is no longer any tension between theory and observation (see also \citealt{claret2010}).

Left unanswered is why both stars have such high obliquities. A tertiary companion star has been invoked to explain the large obliquities in DI Herculis \citep{albrecht2009,philippov2013}. However, the required properties of the tertiary companion have not been thoroughly investigated. The companion must induce strong gravitational perturbations while simultaneously avoiding detection based on existing radial velocity and imaging data. \cite{anderson2017} studied the general problem of obliquity excitation in hierarchical triples due to secular perturbations. If the inner and outer orbits of the triple system are mutually inclined, the outer companion drives nodal precession of the orbital axis of the inner binary. Meanwhile, the oblate host stars precess around the inner orbital axis. Through numerical integrations, \cite{anderson2017} established the criterion for obliquity excitation in hierarchical stellar triples: the spin axis precession frequency must be less than several times the nodal precession frequency of the orbital axis (see Section \ref{sec:misalign} for further details). 
Only if this criterion is satisfied can the spins become decoupled from the orbit, allowing obliquities to be excited to large values. 
The criterion is valid even for binaries undergoing large-amplitude Lidov-Kozai eccentricity/inclination cycles \citep{lidov1962,kozai1962}. 

Regarding DI Herculis in particular, \cite{anderson2017} pointed out that a tertiary companion would need to be located within several AU to have excited the obliquities. However, they did not explore the implied observable consequences of such a close companion. In particular, at such close distances, a companion would contribute to the apsidal precession of the binary, possibly rendering the net apsidal motion rate inconsistent with the observed value. Such a close tertiary companion might also induce large changes in the eccentricity and inclination of the binary, which would be manifested as variations in eclipse times and durations. The implied orbital variations might even prevent the binary from remaining in an eclipsing configuration over the timespan of historical observations.

In this paper, we revisit the hypothesis that secular perturbations from an unseen tertiary companion are responsible for the high obliquities in DI Herculis. We explore whether any companion capable of exciting the obliquities
could have evaded detection thus far. We consider the possibility that the obliquity excitation has occurred at the present orbital separation of DI Herculis. We also consider the possibility that the orbit of DI Herculis was initially larger, and subsequently shrank due to tidal high-eccentricity migration \citep[e.g.][]{fabrycky2007,naoz2014}, causing the obliquities to be excited in the process of migration. The former case requires a close tertiary companion to be present, while the latter case allows for a much more distant companion. We also take advantage of a recent determination of the stars' spin periods ($\approx$1~day) based on quasiperiodic photometric variability \citep{liang2022}. Using the spin periods, we update the calculation of the expected apsidal motion rate in the absence of a tertiary companion, and demonstrate that the theoretical prediction remains consistent with the apsidal motion rate derived by \cite{claret2010}.

Below, in Section \ref{sec:setup}, we describe the secular evolution of suitable hierarchical triples and review the requirements for the outer body to misalign the spins of the inner binary. In Section \ref{sec:results}, we sample over a wide range of tertiary properties and orientations, and evaluate the dynamical effects of the tertiary on the inner binary, assuming that the inner binary has the same physical and orbital properties as are currently observed. In Section \ref{sec:highemig}, we consider obliquity excitation in combination with high-eccentricity tidal migration. We conclude in Section \ref{sec:conclusion}.

\section{Setup \& Dynamical Evolution}
\label{sec:setup}
\subsection{Equations of Motion \& Precession Rates}
\label{sec:equations}

We consider a stellar binary with masses $m_1$ and $m_2$ (the ``inner binary''), and a third body with mass $m_3$ (the ``perturber'') orbiting around the center of mass of $m_1$ and $m_2$ (the ``outer binary''). The orbit of the inner binary is characterized by its angular momentum vector $\Lvecfull$ and eccentricity vector $\evecfull$; likewise, the orbit of the outer binary is characterized by $\Lvecfull_{\p}$ and $\evecfull_{\p}$ (the subscript ``\p'' stands for perturber). The stars of the inner binary have radii $R_1$ and $R_2$, angular rotation frequencies $\Omega_{\star,1}$ and $\Omega_{\star,2}$, and spin angular momentum vectors $\Svecfull_1$ and $\Svecfull_2$.
It will be useful to define
$m_{12} = m_1 + m_2$, and
to refer to the unit vectors $\lvec = \Lvecfull / L$, $\lpvec = \Lvecfull_{\p} / L_\p$, $\svec_1 = \Svecfull_1/S_1$, and $\svec_2 = \Svecfull_2/S_2$.

The orbit of the inner binary evolves due to secular perturbations from $m_3$, along with apsidal precession due to short-range forces (GR, tides, and rotational distortion) and nodal precession due to each stellar quadrupole\footnote{We neglect any orbital evolution due to dissipative tides. The large eccentricity and high obliquities that are observed suggest that tidal dissipation has not been significant during the binary's relatively short lifetime (although see Section \ref{sec:highemig}).}. The secular equations of motion for the inner orbit are therefore
\ba
\frac{\D \Lvecfull}{\D t} & = & \bigg(\frac{\D \Lvecfull}{\D t} \bigg)_{\rm per} + \sum_{i = 1}^{2} \bigg(\frac{\D \Lvecfull}{\D t} \bigg)_{\rm rot,i} \\ 
\frac{\D \evecfull}{\D t} & = &\bigg(\frac{\D \evecfull}{\D t} \bigg)_{\rm per} + \bigg(\frac{\D \evecfull}{\D t} \bigg)_{\rm gr} \\
& + & \sum_{i = 1}^{2} \bigg[ \bigg( \frac{\D \evecfull}{\D t} \bigg)_{{\rm tide},i} + \bigg(\frac{\D \evecfull}{\D t} \bigg)_{{\rm rot},i} \bigg]. \nonumber
\ea

To quadrupole order in the semi-major axis ratio $a/a_{\p}$, the secular equations of motion of the inner binary due to the perturber are \cite[e.g.][]{liu2015},
\ba
\bigg(\frac{\D \Lvecfull}{\D t} \bigg)_{\rm per} & = & \frac{3 L}{4 \tsec j} \bigg[j^2 (\lvec \cdot \lpvec) \lvec \times \lpvec - 5 (\evec \cdot \lpvec) \evec \times \lpvec \bigg], \label{eq:LK_L} \\ 
\bigg(\frac{\D \evecfull}{\D t} \bigg)_{\rm per} & = & \frac{3 j}{4 \tsec} \bigg[(\lvec \cdot \lpvec) \evec \times \lpvec + 2 \lvec \times \evec - 5 (\evec \cdot \lpvec) \lvec \times \lpvec \bigg], \nonumber \\
\label{eq:LK_e}
\ea
where $j = \sqrt{1 - e^2}$. The characteristic timescale for secular perturbations is
\be
\tsec = \frac{m_{12}}{m_3} \frac{\apereff^3}{a^3} n^{-1},
\label{eq:tsec}
\ee
where $n = 2 \pi / P$ is the orbital mean motion of the inner binary, and $\apereff = \aper \sqrt{1 - \eper^2}$ is the effective semi-major axis. In this section of the paper, we neglect the octupole terms because they vanish for equal-mass binaries and the stars of DI Herculis have nearly equal mass. The outer orbit also evolves due to torques from the inner binary. However, our goal in this section is to derive instantaneous quantities for the inner binary (e.g. the apsidal precession rate). As a result, the equations of motion for the outer binary are not relevant for the purposes of this discussion. For sufficiently high mutual inclinations, equations (\ref{eq:LK_L}) and (\ref{eq:LK_e}) lead to Lidov-Kozai cycles \citep{lidov1962,kozai1962}, in which the eccentricity and inclination of the inner binary undergo large variations.

The eccentricity vector precesses due to GR, tides, and rotational distortion. The contributions from GR and tides are
\be
\bigg(\frac{d \evec}{d t} \bigg)_{\rm gr} = \dot{\omega}_{\rm gr} (\lvec \times \evec),
\ee
and (with $i = 1,2$)
\be
\bigg(\frac{d \evec}{d t} \bigg)_{{\rm tide},i} = \dot{\omega}_{{\rm tide},i} (\lvec \times \evec),
\ee
where
\be
\dot{\omega}_{\rm gr} = \frac{3 G m_{12}}{c^2 a(1-e^2)} n,
\ee
and
\be
\dot{\omega}_{{\rm tide},i} = 15 \kappa_i \frac{m_{i'}}{m_i} \frac{R_i^5}{a^5} \frac{1 + \frac{3}{2} e^2 + \frac{1}{8} e^4}{(1 - e^2)^5} n,
\label{eq:omegatide}
\ee
where $\kappa_i$, the apsidal motion constant, depends on the
density distribution of the star, and where the index $i' \neq i$ indicates the opposite star.

Each of the oblate stars in the inner binary experiences a torque from the other star, causing mutual precession of the spin and orbital axes:
\ba
\bigg(\frac{d \Svecfull_i}{d t} \bigg)_{\rm rot} & = & \Omega_{S,i} \cos \theta_i (\Svecfull_i \times \lvec) \\
\bigg(\frac{d \Lvecfull}{d t} \bigg)_{\rm rot} & = & \sum_{i = 1}^{2} \frac{S_i}{L} \Omega_{S,i} \cos \theta_i (\Lvecfull \times \svec_i)
\ea
where $\theta_i$ are the obliquities and the precession frequency is
\be
\Omega_{S,i} = \frac{\kappa_i}{k_{\star,i}} \frac{m_{i'}}{m_i} \frac{R_i^3}{a^3 j^3} \Omega_{\star,i}.
\label{eq:omegarot}
\ee
In equation (\ref{eq:omegarot}), $k_{\star,i} = S_i / m_i R_i^2 \Omega_{\star,i}$ is the dimensionless moment-of-inertia constant.

Each stellar quadrupole causes the eccentricity vector to precess according to 
\be
\bigg(\frac{d \evec}{d t} \bigg)_{{\rm rot},i} = \Omega_{e,i} \bigg[\frac{1}{2} (5 \cos^2 \theta_i - 1) (\lvec \times \evec) - \cos \theta_i (\svec_i \times \evec) \bigg],
\label{eq:precession_rot_full}
\ee
\be
\Omega_{e,i} = \kappa_i \bigg(\frac{R_{i}}{a} \bigg)^2 \frac{\hat{\Omega}_{\star,i}^2}{(1 - e^2)^2} n,
\ee
where $\hat{\Omega}_{\star,i} = \Omega_{\star,i} / \sqrt{G m_i/R_i^3}$ is the spin frequency in units of the breakup frequency.

We calculate the net apsidal precession rate directly from the vector equations of motion. We adopt the procedure described in Appendix B of \cite{philippov2013}, generalized to include the dynamical effects of the tertiary. We construct an observer-oriented coordinate system $(\xvec,\yvec,\zvec)$, with the $\zvec$-axis along the line of sight and the $\xvec-\yvec$ plane in the sky plane. Without loss of generality, we place $\lvec$ in the $\xvec$-$\zvec$ plane, inclined by an angle $I$ relative to $\zvec$. With these choices,
\be
\lvec = \sin I\,\xvec + \cos I\,\zvec.
\ee
In this coordinate system, the $\yvec$ axis is the line of nodes,
which varies in direction on secular timescales, and the eccentricity unit vector of the inner binary is
\be
\ehatvec = \cos \omega\,\yvec + \sin \omega\,(\lvec \times \yvec),
\ee
with $\omega$ denoting the argument of pericenter of the inner orbit. The outer orbit is specified according to its inclination $I_{\p}$, longitude of ascending node $\Omega_{\p}$, and argument of pericenter $\omega_{\p}$.

In the $(\xvec,\yvec,\zvec)$ coordinate system, each spin axis $\svec_i$ has the spherical coordinates
\be
\svec_i = \sin I_{\star,i} \cos \lambda_i\,\xvec + \sin I_{\star,i} \sin \lambda_i\,\yvec + \cos I_{\star,i}\,\zvec,
\ee
with $I_{\star,i}$ the inclination of the stellar equator relative the line of sight, and $\lambda_i$ the sky-projected obliquity. Upon specification of $I$, $I_{\star,i}$ and $\lambda_i$, the true obliquity $\theta_i$ can be calculated using the equation
\be
\cos \theta_i = \sin I \sin I_{\star,i}  \cos \lambda_i + \cos I \cos I_{\star,i}.
\label{eq:trueobliquity}
\ee

The net apsidal precession rate of the inner orbit ($\dot{\omega}_{\rm}$) is calculated according to
\be
\dot{\omega} = -\frac{1}{\sin \omega} \bigg(\frac{\D \ehatvec}{\D t} \cdot \yvec + \frac{\D \yvec}{\D t} \cdot \ehatvec \bigg),
\ee
with
\be
\frac{\D \yvec}{\D t} = \frac{1}{\sin^2 I} \bigg[\sin I \zvec \times \frac{\D \lvec}{\D t} + \cot I \bigg(\zvec \cdot \frac{\D \lvec}{\D t} \bigg) (\zvec \times \lvec) \bigg],
\label{eq:dydt}
\ee
and
\ba
\frac{\D \ehatvec}{\D t} & = & \frac{1}{e} \bigg[\frac{\D \evecfull}{\D t} - \bigg(\ehatvec \cdot \frac{\D \evecfull}{\D t} \bigg) \ehatvec \bigg] \\
\frac{\D \lvec}{\D t} & = & \frac{1}{L} \bigg[\frac{\D \Lvecfull}{\D t} - \bigg(\lvec \cdot \frac{\D \Lvecfull}{\D t} \bigg) \lvec \bigg].
\ea
The expression for $\D \yvec / \D t$ in equation (\ref{eq:dydt}) includes the contribution from the tertiary companion and from each of the oblate stars.

\subsection{Requirements for Generating Spin-Orbit Misalignment}
\label{sec:misalign}
In order for an inclined tertiary to produce spin-orbit misalignment in the inner binary, the secular perturbation from the tertiary must be sufficiently strong compared to the spin-orbit coupling between the inner binary members \citep{anderson2017}. Even if the inner binary is undergoing large amplitude Lidov-Kozai cycles (so that the orbital inclination of the inner binary undergoes dramatic oscillations), spin-orbit misalignment in star $i$ may only be generated if the orbital axis precession frequency $\Omega_{L} \sim 1/t_{\rm sec}$ (equation \ref{eq:tsec}) is comparable to or exceeds the spin-axis precession frequency $\Omega_{S,i}$ (equation \ref{eq:omegarot}).

Through an ensemble of numerical integrations of inclined hierarchical triples with a wide range of properties, \cite{anderson2017} determined that spin-orbit misalignment of star $i$ can be essentially guaranteed if
\be
\A_i = \bigg(\frac{\Omega_{S,i}}{\Omega_L} \bigg)_{e = 0} \lesssim 3.
\label{eq:A}
\ee
Conversely, if the precession rate of $\svec$ around $\lvec$ is roughly three times larger than the precession rate of $\lvec$ around $\lpvec$, then large spin-orbit misalignment is unlikely to be possible. This criterion was shown to hold even for systems undergoing extreme eccentricity and inclination oscillations.

The criterion for obliquity excitation proposed by \cite{anderson2017} is based upon an expansion of the disturbing potential of the tertiary to quadrupole-order in the semi-major axis ratio, and was numerically confirmed to be valid even when the octupole terms contribute. The referee of this paper raised the concern that higher-order terms may significantly alter the dynamical evolution. In particular, previous work has argued that under some circumstances, the hexadecapole terms may cause the orbital inclination to flip from prograde to retrograde values \citep{will2017}. However, for DI Herculis, we think the quadrupole-order truncation is appropriate for the following reasons. In order for the hexadecapole terms to be important, they must be sufficiently large compared to other effects, such as perturbations due to general relativity, tides, and rotational oblateness. Since the hexadecapole terms scale as $(m_3/m_{12}) (a/a_{\rm p})^5$, and $a/a_{\rm p}$ must be a small parameter for the system to remain hierarchical, one would need a very massive tertiary for the hexadecapole terms to alter the system dynamics. For the case of DI Herculis, there is unlikely to be a realistic volume in parameter space where the hexadecapole terms are important, because massive tertiaries are ruled out by observations, and because the system must satisfy dynamical stability. Furthermore, even if the hexadecapole terms were to contribute meaningfully to the orbital evolution, the obliquity excitation criterion should still depend primarily on the quadrupole-order precession rate.

To confirm these expectations, we performed some numerical experiments on stellar triples using the GRIT N-body package \citep{chen2021}\footnote{https://github.com/GRIT-RBSim/GRIT}, which evolves obliquities simultaneously with the orbits. 
We set the masses, radii, spin periods, and initial orbital parameters of the inner binary
to be the measured values for DI Herculis, and adopted a nominal tertiary mass of 1 solar mass. We set the initial mutual inclination between the inner and outer orbits to 85 degrees and the outer eccentricity to $0.2$,
and varied the outer semi-major axes between a minimum of about 0.75 AU
and a maximum of about 3 AU. Varying the outer semi-major axis in this range causes the spin-orbit coupling parameters $\A_i$ to vary between about 0.5 and 8. The large mutual inclination leads to large-amplitude Lidov-Kozai cycles and large inclination variations. We integrated each system for 1000 secular timescales, and recorded the maximum value of the obliquity. These experiments confirmed that in spite of large inclination variation, obliquities are not excited beyond a few degrees
in systems with adiabaticity parameters $\A_i$ exceeding $3$-$4$. This is in agreement with the results obtained by \cite{anderson2017} using a secular octupole-order code.

We note that equation (\ref{eq:A}) assumes a circular orbit, while DI Herculis has an eccentricity of about $0.5$. Including the eccentricity dependence in the precession frequencies would introduce an order-unity correction to equation (\ref{eq:A}). Since the eccentricity may oscillate due to secular perturbations, we neglect such order-unity corrections in the definition of $\mathcal{A}_i$.

For an inner binary with equal masses and radii, the requirement $\A \lesssim 3$ implies that the perturber semi-major axis, eccentricity, and mass must satisfy
\ba
\frac{a_{\rm p,eff}}{\bar{m}_3^{1/3}} & \lesssim & 2 \, {\rm AU} \bigg(\frac{m_{12}}{10 \msun} \bigg)^{1/3} \bigg(\frac{R_{\star}}{2.5 \rsun} \bigg)^{-1} \bigg(\frac{P_{\star}}{1 {\rm \, d}} \bigg)^{1/3} \nonumber \\
&& \times \bigg(\frac{P_{\rm orb}}{10 \, {\rm d}} \bigg),
\label{eq:amax}
\ea
where $\bar{m}_3 \equiv m_3 / \msun$.
Thus, for a DI-Herculis-type binary, a tertiary on a circular orbit must be located within $\sim$2~AU to have excited the obliquity in the present orbital configuration. 

On the other hand, the hierarchical triple must be dynamically stable. We adopt the semi-analytic stability criterion from \cite{mardling2001}:
\be
\frac{\aper}{a} \gtrsim 2.8 \bigg(1 + \frac{m_3}{m_{12}} \bigg)^{2/5} \frac{(1 + e_{\rm p})^{2/5}}{(1 - e_{\rm p})^{6/5}} \bigg[1 - 0.3 \frac{I_{\rm mut}}{\pi} \bigg],
\label{eq:stability}
\ee
where $I_{\rm mut}$ is the inclination between the inner and outer orbits. This stability criterion has been tested numerically by \cite{he2018}, and was found to perform well for stellar triples.

In order for such a tertiary companion to have evaded radial-velocity (RV) detection thus far, we require the semi-amplitude $K$ to be smaller than a critical value $K_{\rm crit}$. The requirement that the radial velocity semi-amplitude not exceed a critical threshold $K_{\rm crit}$ translates into the condition
\ba
\aper (1 - \eper^2)^2 & \gtrsim & 0.89 {\rm AU} \bigg(\frac{K_{\rm crit}}{10 {\ \rm km \ s^{-1}}} \bigg)^{-2} \bigg(\frac{m_{12} + m_3}{10 \msun} \bigg)^{-1} \nonumber \\
&& \times \bigg(\frac{m_{3} \sin I_{\p}}{\msun} \bigg)^2.
\ea
We chose $K_{\rm crit} = 10\,$ km\,s$^{-1}$, based on visual inspection of the radial-velocity curves and residuals presented by \cite{albrecht2009}. Larger values of $K_{\rm crit}$ would be compatible with the data only if the orbital period is several decades or longer. However, there is no need to consider such long period tertiaries because they would be unable to excite the obliquities of the inner binary.

Finally, in order for a tertiary companion to have evaded detection via direct imaging, we require the luminosity of the tertiary companion to be below a certain fraction of the binary luminosity. Based on the light-curves, \cite{martynov1980} found that the luminosity of the tertiary must satisfy $L_3 < 0.03 (L_1 + L_2)$. In terms of mass, this condition translates into $m_3 \lesssim 2.5 \msun$ \citep{guinan1985}.

\begin{figure*}
\centering 
\includegraphics[width=0.9\textwidth]{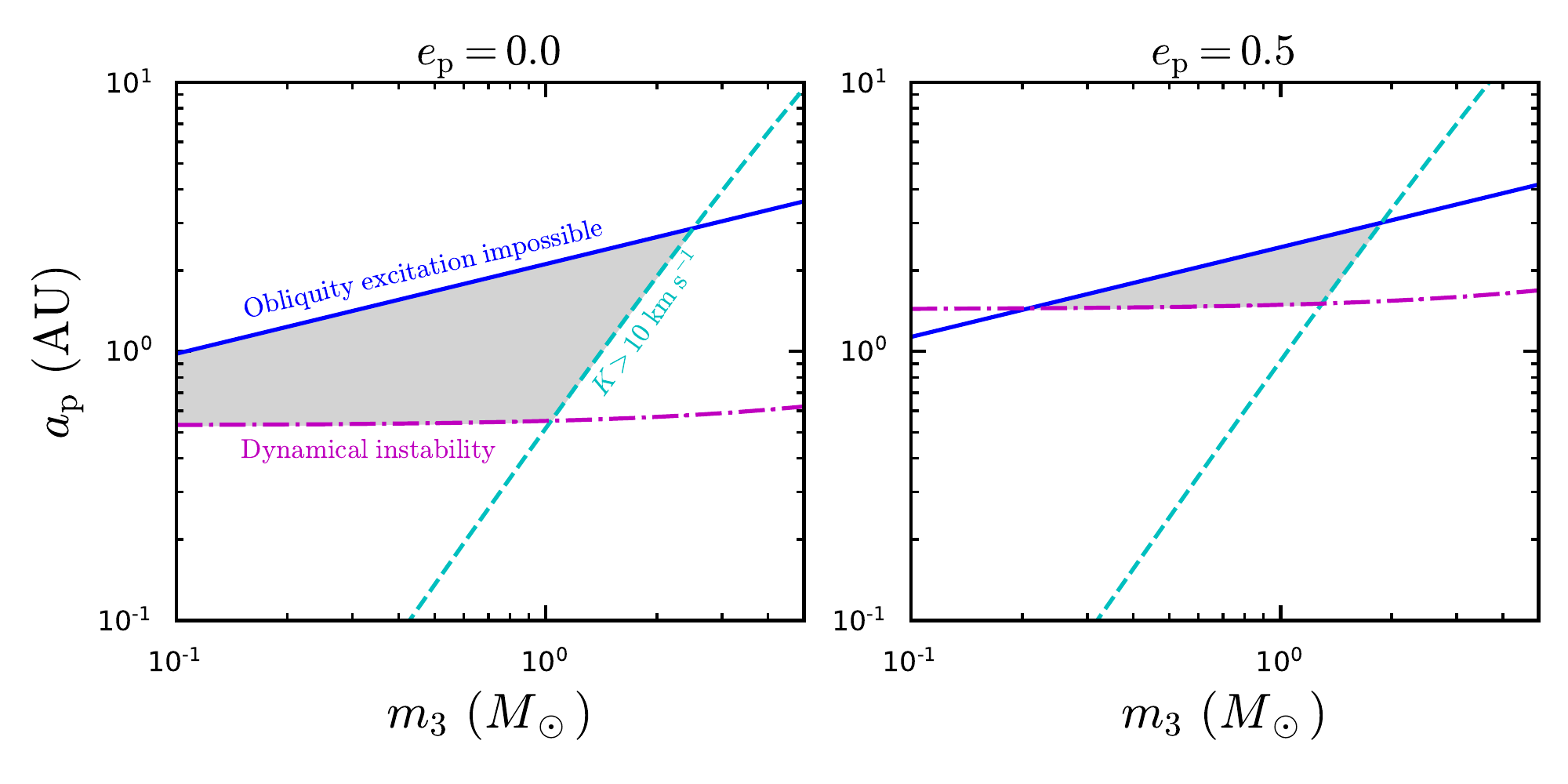}
\caption{Parameter space for an unseen tertiary body. With parameters inside the gray shaded region, a tertiary might be capable of exciting the obliquities in DI Herculis while simultaneously satisfying dynamical stability and evading RV detection. The solid blue line is based on equation (\ref{eq:amax}); perturbers
with properties above this line are not strong enough to excite the obliquities of the DI Herculis members. The dashed cyan line shows $K = 10$ km/s (assuming $\sin I_{\rm p} = 0.8$). The magenta dot-dashed line depicts the dynamical stability condition from \cite{mardling2001}, assuming a mutual inclination of $40^\circ$. The left panel is for the case of a circular tertiary orbit, and the right panel is for an orbit with $\eper = 0.5$.}
\label{fig:paramspace}
\end{figure*}

Figure \ref{fig:paramspace} depicts the allowed region in the space of mass and separation for an unseen tertiary to have excited the obliquity in DI Herculis, while satisfying the requirement for long-term dynamical stability and evading RV detection.

\section{Obliquity Excitation at the Present Orbital Separation}
\label{sec:results}

In this section we sample a wide range of properties for the tertiary companion and evaluate the predicted apsidal precession rate, along with other potentially observational quantities. We fix the parameters of DI Herculis at the values given in Table \ref{table}.

\begin{deluxetable}{lll}
\tablecaption{Adopted properties of DI Herculis. \label{table}}
\tabletypesize{\small}
\startdata
\tablehead{\colhead{Quantity} & \colhead{Value} & \colhead{Reference}}%
Orbital period & $10.55 \, {\rm days}$ & 2 \\
Eccentricity & 0.48 & 1 \\
Inclination & $89.3^\circ$ & 1 \\
Primary mass & $5.2 \msun$ & 1 \\
Secondary mass & $4.6 \msun$ & 1 \\
Primary radius & $2.7 \rsun$ & 1 \\
Secondary radius & $2.5 \rsun$ & 1 \\
Primary $v \sin I$ & $108 \, {\rm km \ s^{-1}}$ & 3 \\
Secondary $v \sin I$ & $116 \, {\rm km \ s^{-1}}$ & 3 \\
Apsidal motion constant $\kappa$ & 0.008 & 4 \\
Moment of inertia constant $k_{\star}$ & 0.06 & 4 \\
\enddata
\tablerefs{(1) \cite{popper1982}, (2) \cite{guinan1985}, (3) \cite{albrecht2009}, (4) \cite{claret2019}}
\end{deluxetable}

The dynamical evolution and apsidal precession rate depend on the spin periods of the inner binary members. We consider two different cases assuming different levels of knowledge about the spin periods: 
\begin{enumerate}
\item We assume the inclination $I_\star$ of each spin axis relative to the line of sight is drawn from an isotropic distribution (uniform in
$\cos I_\star$). Then, we calculate the spin period based on the observed $v \sin I_\star$, stellar radius, and inclination, adopting a $10 \%$ uncertainty in $v \sin I_\star$. This sampling procedure results in broad distributions for the spin periods and true obliquities (as implied from equation \ref{eq:trueobliquity}). We refer to this sampling procedure as \texttt{Unknown Period}.

\item We assume the spin period of each star is $1.0\pm 0.1$ day. This assumption is based upon
a recent analysis of DI Herculis using data from the Transiting Exoplanet
Survey Satellite \citep{liang2022}. Based on the out-of-eclipse photometric variability and the effects of gravity darkening on the shape of the eclipse light curves, they concluded that both stars have rotation periods of 1 day, with systematic uncertainties of about 0.1 day. Using an MCMC method, we combined the constraints on the rotation period, radius, and projected rotation velocity for each star to obtain the posterior probability density for $I_\star$ for each star \citep{masuda2020}.  This, in turn, narrows down the possible values of the true obliquities. We refer to this sampling procedure as \texttt{Known Period}.
\end{enumerate}

\begin{figure*}
\centering 
\includegraphics[width=0.9\textwidth]{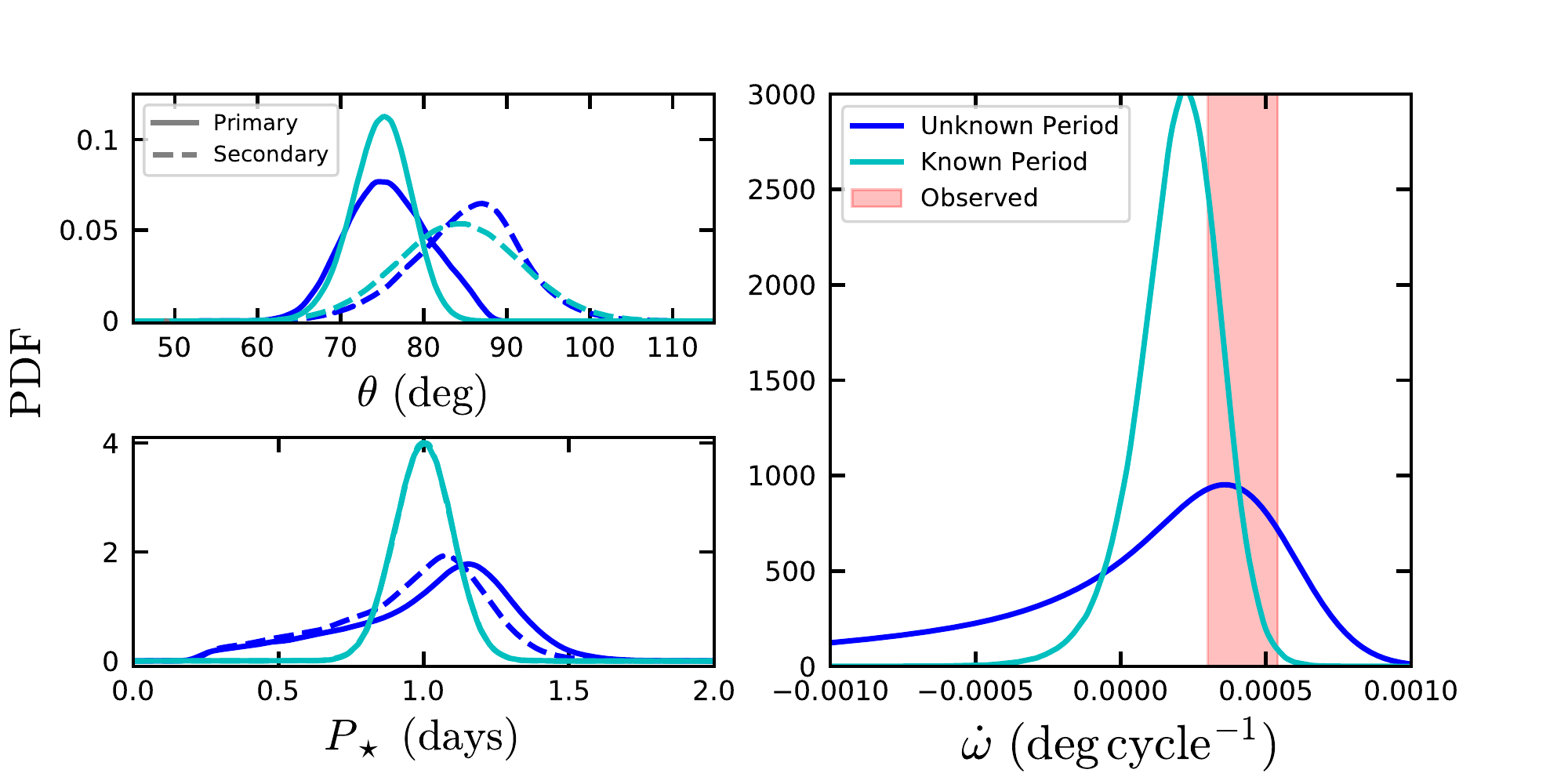}
\caption{Distributions (obtained from a Gaussian kernel density estimate) of true obliquities (top left), spin periods (bottom left), and apsidal precession rates (right) for the case of no tertiary companion. The blue curves show the \texttt{Unknown Period} sampling procedure, the cyan curves show the \texttt{Known Period} procedure, and the solid and dashed curves in the left panels denote the primary and secondary respectively. The shaded red region depicts the observed apsidal precession rate reported by \cite{claret2010}.  }
\label{fig:notertiary}
\end{figure*}

In both cases, we fix the masses, radii, semi-major axis, orbital inclination, eccentricity, and argument of pericenter of DI Herculis at the values given in Table 1. To start, we consider the case of no tertiary companion (so that the contributions to the apsidal precession rate consist of GR, tidal, and rotational deformation alone). Figure \ref{fig:notertiary} shows the resulting distributions of true obliquities, spin periods, and apsidal precession rates obtained from the \texttt{Unknown Period} and \texttt{Known Period} sampling procedures. The \texttt{Unknown Period} sampling procedure reproduces the obliquity distribution obtained by \cite{claret2010} under the same assumptions. The \texttt{Known Period} sampling procedure results in a narrower distribution of apsidal precession rates that remain consistent with the observed rate obtained by \cite{claret2010}.

\begin{figure*}
\centering 
\includegraphics[width=0.9\textwidth]{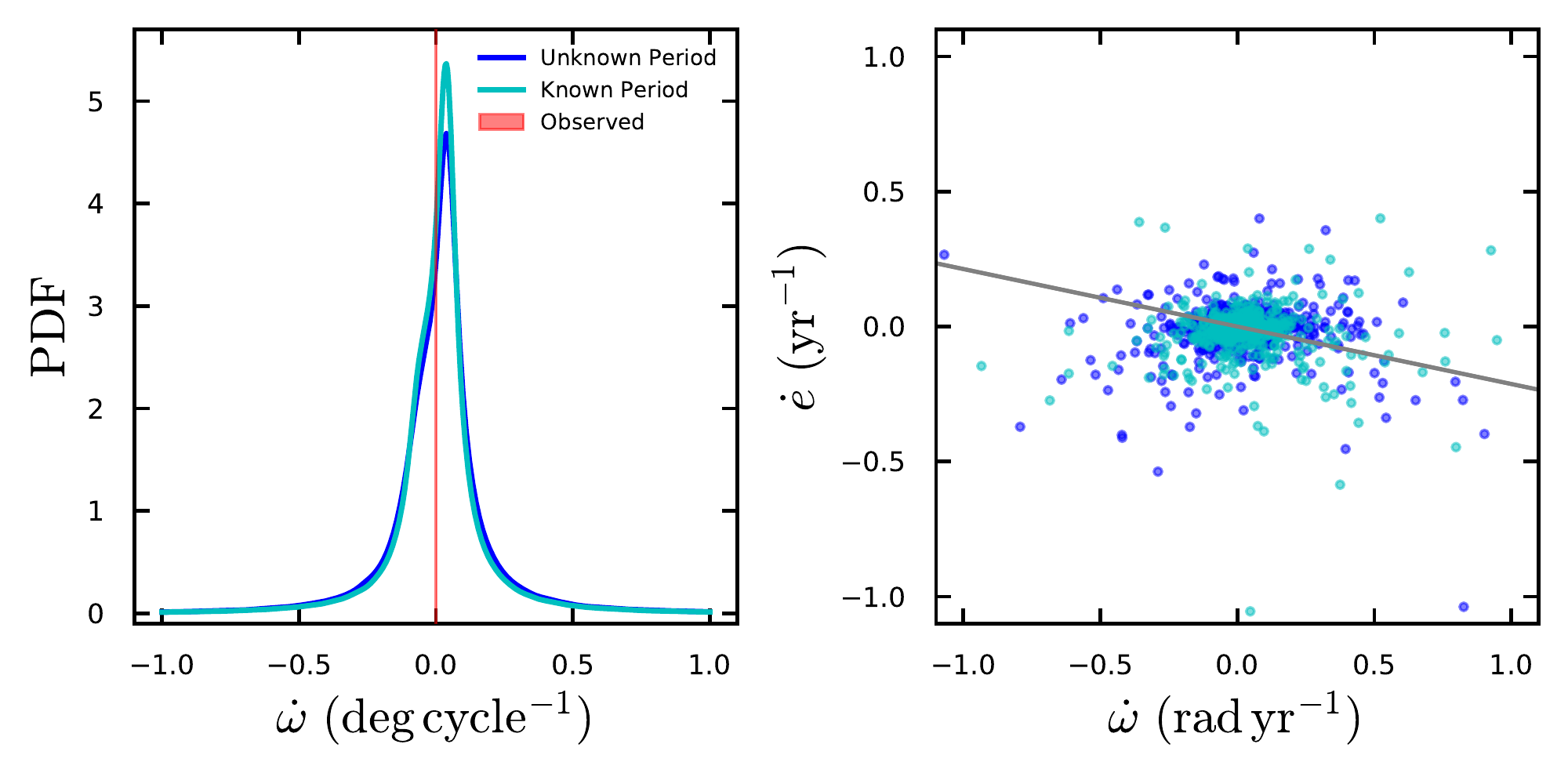}
\caption{{\it Left:} Similar to the right panel of Fig.~\ref{fig:notertiary}, we show the distributions of the apsidal precession rate for the \texttt{Unknown Period} sampling procedure (blue), and the \texttt{Known Period} procedure (cyan). The red line depicts the observed apsidal precession rate obtained by \cite{claret2010} (the width of the line indicates the observational uncertainty). {\it Right:} Instantaneous $\dot{e}$ versus $\dot{\omega}$ for the sampled systems. Note that for visual clarity, here we plot only 1000 randomly selected points from each sampling procedure. The gray line indicates the combinations of $\dot{\omega}$ and $\dot{e}$ that are compatible with observations.}
\label{fig:withtertiary}
\end{figure*}

Next we add a tertiary with properties determined as follows. We sample the semi-major axis uniformly between 0.6 and 10~AU, and the eccentricity uniformly between 0 and 0.9. The lower limit on the semi-major axis corresponds to $\aper = 3 a$. The perturber mass is sampled uniformly between 0.1 and 2.5\,$\msun$, where the upper limit is chosen so that the tertiary could have evaded detection through imaging \citep{martynov1980}. We sample $\cos I_{\rm p}$ from a uniform distribution between $-1$ and 1 (as appropriate for an isotropic distribution of orientations), and we sample the longitude of ascending node $\Omega_{\rm p}$ and argument of pericenter $\omega_{\rm p}$ from uniform distributions between $0$ and $2\pi$.
We reject systems for which any of the following statements is true:
\begin{itemize}
   \item $\A_i > 3$ (see equation \ref{eq:A}). Such configurations do not result in significant obliquity excitation, due to the strong torques coupling the spins with the inner orbit.
    
    \item The system is dynamically unstable according to the \cite{mardling2001} criterion (see equation \ref{eq:stability}).
    
    \item The implied RV semi-amplitude of the center of mass of the inner binary exceeds $10$ ${\rm km \, s^{-1}}$. As discussed previously, such tertiaries are likely to have been detected in RV studies.
    
\end{itemize}
After sampling a large number of total systems and applying the aforementioned cuts, we are left with roughly $5 \times 10^5$ systems for both the \texttt{Unknown Period} and \texttt{Known Period} sampling procedures. The left panel of Fig.~\ref{fig:withtertiary} shows the distribution of instantaneous net apsidal precession rates (accounting for the contributions from all short-range forces and the perturber). Due to the wide range of properties we sampled for the perturber, and the strong dependence of $\dot{\omega}$ on the relative orientations of the inner and outer orbits, the distribution of apsidal precession rates is broad and spans both positive and negative values. The observed $\dot{\omega}$ is much closer to zero than the breadth of the theoretical distribution. Thus, a high degree of fine tuning would be required for a perturber to avoid producing too much apsidal precession. The percentage of sampled systems that produce slow enough apsidal precession to be compatible with the observed value is $\sim$0.1\%.

A subtlety in this comparison arises because $\dot{\omega}$ is not a directly measured quantity. Instead, the ``observed'' value of $\dot{\omega}$ is obtained by fitting a parametric model to a collection of
mid-eclipse times (the times of minimum light). The parametric model
is based on the premise that the only orbital parameter
that varies with time is the argument of pericenter.
However, if a tertiary star is present, the eccentricity and inclination
would also vary with time, invalidating the premise of the model.
Thus, the preceding comparison between the ``observed'' and
calculated values of the apsidal precession rate is not correct
in detail.

To account for this subtlety, we need to allow for the effects
of eccentricity and inclination variations on the observed mid-eclipse times. Ultimately, the constraints on the apsidal precession rate arise from observed changes in the quantity
\be
D = T_{\rm II} - T_{\rm I} - \frac{1}{2} P,
\ee
where $T_{\rm I}$ and $T_{\rm II}$ are the times of minimum light for the primary and secondary eclipses, and $P$ is the orbital period \citep[see][]{guinan1985}. For inclinations close to $90^\circ$, as is currently the case for DI Herculis, a very good approximation for $D$ is \citep{sterne1939}
\be
D = \frac{P}{\pi} \bigg[\tan^{-1}\bigg(\frac{e \cos \omega}{\sqrt{1 - e^2}} \bigg) + \frac{e \cos \omega}{1 - e^2 \sin^2 \omega } \sqrt{1 - e^2} \bigg].
\label{eq:D}
\ee
Differentiating equation (\ref{eq:D}) and using the observed values of $e$ and $\omega$ for DI Herculis, the rate of change of $D$ can be expressed as 
\be
\dot{D} = 1.2~{\rm days}~(\dot{\omega} + 4.7 \dot{e}). 
\label{eq:ddot}
\ee
We see from this equation that apsidal motion and eccentricity variations
both contribute to the observed departures from
periodicity of the eclipses.

The right panel of Fig.~\ref{fig:withtertiary} shows the instantaneous $\dot{e}$ versus $\dot{\omega}$ obtained from our Monte-Carlo sampling procedure (shown as dots), along with the combination of $\dot{\omega}$ and $\dot{e}$ that are compatible with the apsidal precession rate found by \cite{claret2010} (gray line).

Inspection of Fig.~\ref{fig:withtertiary} shows that most of the sampled tertiary companions will result in changes of order unity per year in both $\omega$ and $e$ of the inner binary. These large variations are the result of Lidov-Kozai cycles. Correspondingly large changes in the orbital inclination (relative to the line of sight) are also expected. The large instantaneous values of $\dot{\omega}$, $\dot{e}$, and $\dot{I}$ imply that most of the triple systems that we have sampled can be immediately ruled out based on the observed constancy of the orbital parameters over the past 50 years. For example, Fig.~5 of \cite{reisenberger1989} shows that the orbital inclination has remained constant to within $\sim 0.1^\circ$ since the 1970s.

Another subtlety arises because the secular timescales of these hierarchical triples tend to be comparable to (or shorter than) the interval over which DI Herculis has been observed. Thus, we should not compare the {\it instantaneous} rates of change of the orbital elements; we should instead calculate the total change in the orbital parameters over the timespan of historical observations. We did so by directly integrating the secular equations of motion backward 50 years in time. In modeling the secular evolution, we included the mutual perturbation of the inner and outer binaries up to octupole order, along with the spin-orbit torques and apsidal precession from GR, tides, and rotation (see \citealt{liu2015} for the full, octupole-order equations of motion for the inner and outer binaries). First, we selected all triples with an instantaneous value of $\dot{D}$ that falls within the uncertainty of the observed value (constituting several hundred systems each from the \texttt{Unknown Period} and \texttt{Known Period} samples). We found through this procedure that essentially all of the hypothetical tertiary companions can be ruled out because they predict too much variation in eccentricity or inclination.

\begin{figure}
\centering 
\includegraphics[width=0.4\textwidth]{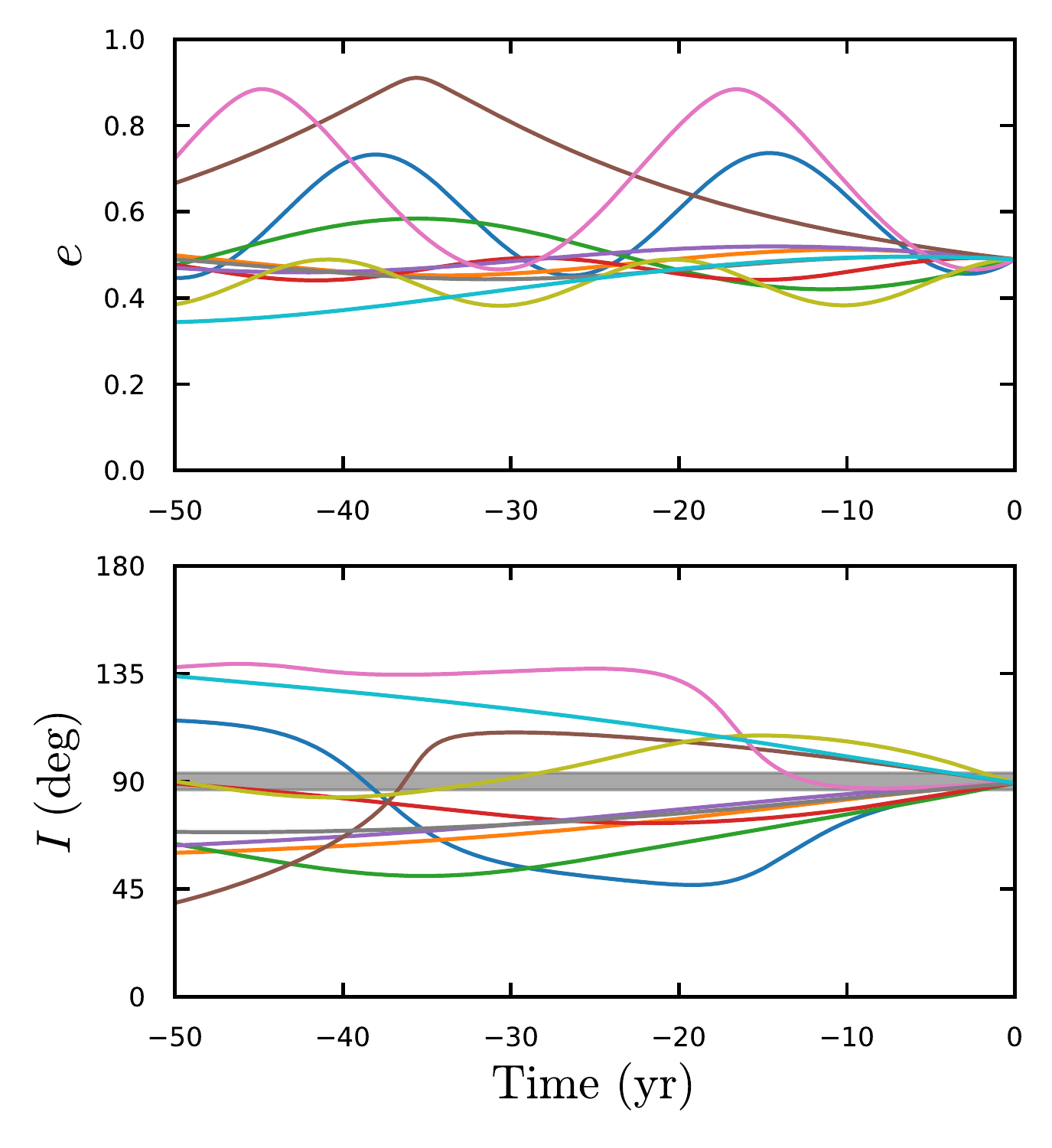}
\caption{{\it Top:} Secular evolution of the inner binary's eccentricity over the past 50 years, starting from the observed orbital state. Plotted are 10 randomly selected triple systems from our sampling procedure for which the instantaneous $\dot{D}$ (see equation \ref{eq:ddot}) is consistent with observations. Due to the compactness of the systems and large mutual inclinations, Lidov-Kozai cycles are induced, leading to large changes in orbital eccentricity and inclination. {\it Bottom:} Evolution of the orbital inclination of the inner binary (relative to the line of sight). The gray shaded regions shows the approximate range of inclinations for which the binary remains transiting.}
\label{fig:timeev}
\end{figure}

\begin{figure}
\centering 
\includegraphics[width=0.4\textwidth]{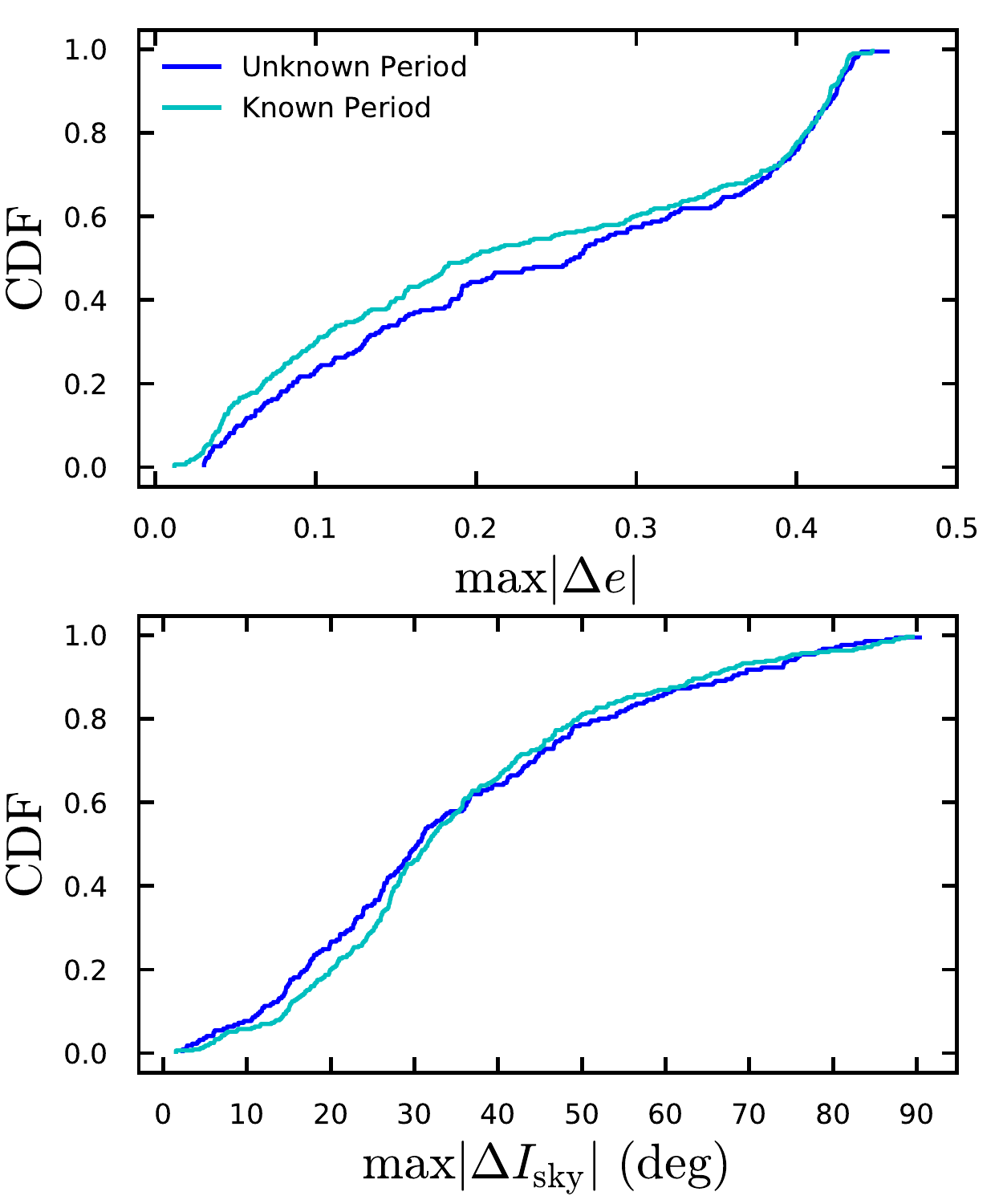}
\caption{Maximum change in the eccentricity (top panel) and orbital inclination (bottom panel) over the previous 50 years, starting from the observed orbital properties of DI Herculis and the sampled configurations for the tertiary companion, and accounting for the secular evolution of the hierarchical triple system.}
\label{fig:delta_e_I}
\end{figure}

Fig.~\ref{fig:timeev} shows the resulting time evolution of 10 randomly selected triple systems having an instantaneous $\dot{D}$ consistent with the observed value. The systems exhibit Lidov-Kozai cycles, which arise from the compactness of the systems and large mutual inclinations and lead to order-unity changes in both the orbital eccentricity and inner binary inclination relative to the line of sight. Such large changes can be immediately ruled out based on the lack of variation of these orbital elements over the observational history of DI Herculis. Fig.~\ref{fig:delta_e_I} shows the cumulative distributions of the maximum changes in eccentricity and orbital inclination of the inner binary over the last 50 years. Over $95\%$ of the systems result in a change in the inclination greater than about $5^\circ$, and $95\%$ result in a change in the eccentricity greater than about $0.05$. In addition to these secular variations of the orbital elements, there would be smaller-amplitude and shorter-timescale variations in the observed eclipse times \citep[e.g.][]{borkovits2015}. We therefore conclude that a tertiary having raised the obliquity while simultaneously avoiding detection is extremely unlikely. 

\section{Obliquity Excitation as a Result of High-Eccentricity Migration}
\label{sec:highemig}

In the preceding section we ruled out the possibility that the high obliquities in DI Herculis are the result of ongoing perturbations from a tertiary companion. However, we have not yet considered the possibility that the orbit of DI Herculis was formerly much wider, and that a tertiary companion was responsible for both the orbital shrinkage and the obliquity excitation. We refer to the frequently-discussed mechanism for forming short-period binaries via high-eccentricity migration due to Lidov-Kozai (LK) cycles induced by a tertiary companion. In this scenario, the eccentricity of the inner binary is periodically excited to an extreme value, causing strong tidal dissipation during pericenter passages, which leads to orbital decay and eventual circularization. 
Such a scenario may seem unlikely because of the observed large eccentricity and high obliquities of DI Herculis, which imply relatively long tidal dissipation timescales compared to the system age. However, at the large eccentricities ($e \sim 1$) needed for migration, dynamical tides may be excited, leading to more efficient tidal dissipation and orbital shrinkage \citep[e.g.][]{moe2018}. The large eccentricity currently observed would imply that the binary is still in the final stages of migration. Previously, \cite{naoz2014} conducted a general population synthesis of close binary formation through Lidov-Kozai cycles, assuming a wide range of stellar masses and initial architectures. Their study produced some close binaries with orbital properties similar to those of DI Herculis. 

Whether high-eccentricity migration may simultaneously reproduce the orbital and spin properties of DI Herculis within the 5 Myr system age, and with a tertiary that would have remained undetected thus far, has not been explored. A high-eccentricity migration origin is harder to study definitively because
it invokes events that took place when the system had very different parameters and involves tidal dissipation timescales that are
very uncertain. Nevertheless, 
in this section we try to assess the plausibility of this scenario. We utilize simple arguments, and leave an in-depth study for future work.

First, we place constraints on the initial architecture of the triple system, working in the quadrupole approximation, and in the limit that the angular momentum of the inner binary is much smaller than the angular momentum of the outer binary. In this limit, the perturber properties enter only through the combination
\be
\label{eq:perturber_parameter}
\frac{\aper^3 \left(1 - \eper^2\right)^{3/2}}{m_3}.
\ee
It is well-known that in order for the tertiary to drive LK cycles in the inner binary, the apsidal precession rate due to short-range-forces (i.e.\ general relativity, tides, and rotational distortion) must be sufficiently slow compared to the apsidal precession rate induced by the tertiary \citep[e.g.][]{fabrycky2007,liu2015,anderson2017}. Note that in some cases, the GR contribution may lead to enhanced eccentricities, even when the GR precession is faster than the precession from the tertiary \citep{naoz2013}. However, whether such eccentricity enhancement is preserved when apsidal precession from tides and rotational distortion are also included is not clear, and we neglect this complication in the following discussion. For an inner binary with the masses and radii of DI Herculis and 1-day rotation periods, the GR precession dominates over the rotational and tidal contributions for $a \gtrsim 0.2$ AU. For this reason, we consider only the role of GR in suppressing LK cycles. For LK cycles of any amplitude to occur in DI Herculis, the outer body must satisfy (see equation 29 of \citealt{anderson2017}) 
\be
\frac{\apereff}{\bar{m}_3^{1/3}} \lesssim 91 \, {\rm AU} \bigg(\frac{a}{{\rm AU}} \bigg)^{4/3} \bigg(1 - \frac{5}{3} \cos^2 I_{\rm mut} \bigg)^{1/3},
\label{eq:migration_condition}
\ee
where $a$ is the initial semi-major axis of DI Herculis.
Equation (\ref{eq:migration_condition}) represents a necessary but not sufficient condition for migration, because only certain mutual inclinations between the inner and outer orbits lead to pericenter distances that are small enough for strong tidal dissipation and inward migration.
In addition, the LK timescale must be shorter than the system age ($\sim 5$ Myr). This implies that the outer body should satisfy
\be
\frac{\apereff}{\bar{m}_3^{1/3}} \lesssim 215 \, {\rm AU} \bigg(\frac{t_{\rm age}}{5 \, {\rm Myr}} \bigg)^{1/3} \bigg( \frac{a}{{\rm AU}}\bigg)^{1/2}.
\label{eq:age}
\ee

Even when there are large variations of the inner binary's inclination, LK-induced high-eccentricity migration does not necessarily result in simultaneous obliquity excitation, due to the strong torques that couple the spins with the inner orbit. Using again the empirical finding by \cite{anderson2017} for obliquity growth (see Section \ref{sec:setup} and equation \ref{eq:A}), the outer body must satisfy equation (\ref{eq:amax}). Finally, the initial configuration of the triple system must be dynamically stable; again, we use equation (\ref{eq:stability})
to decide whether a hypothetical triple satisfies this criterion.

\begin{figure}
\centering 
\includegraphics[width=0.45\textwidth]{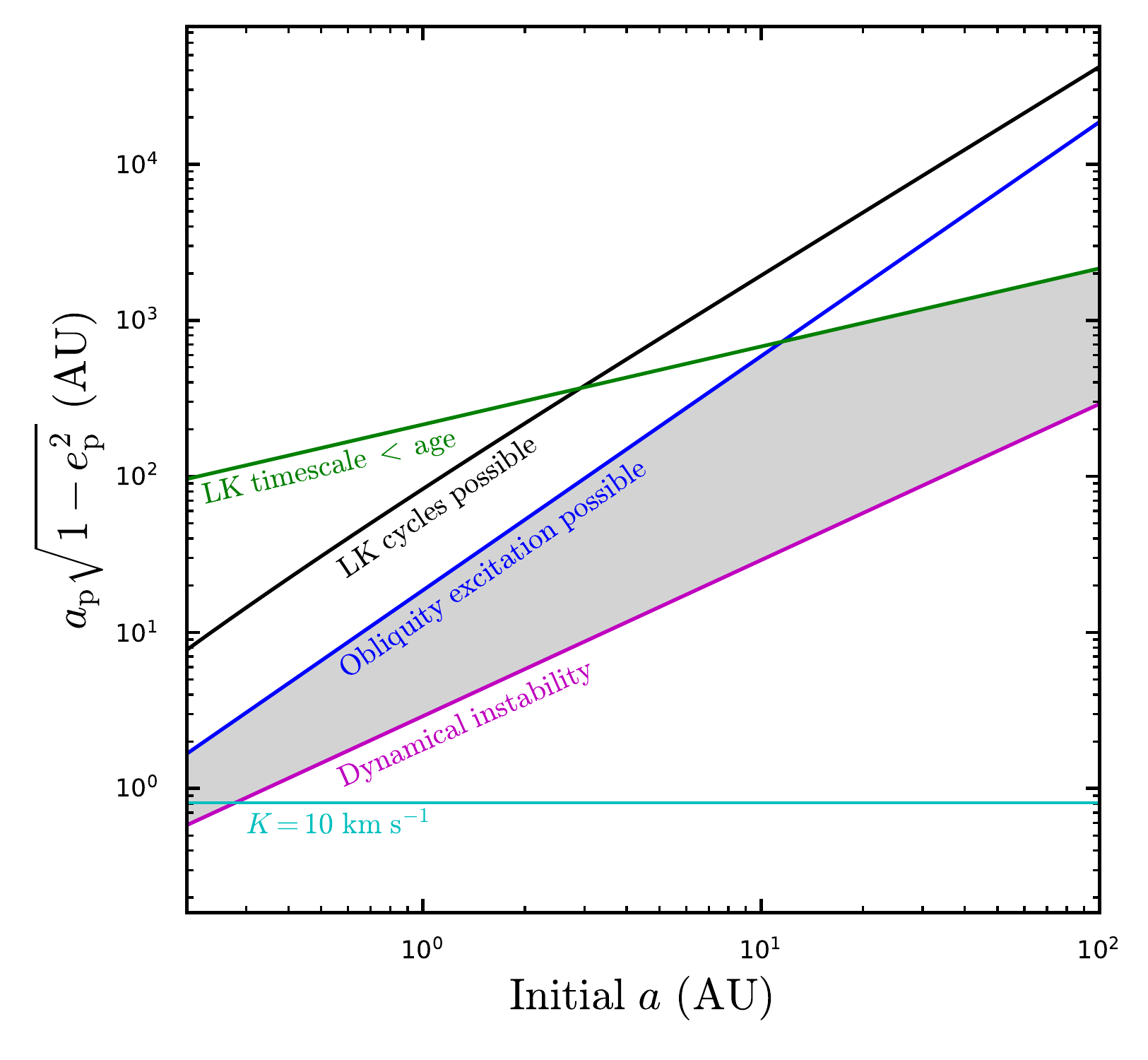}
\caption{Possible parameter space for high-eccentricity migration and simultaneous obliquity excitation driven by a tertiary companion (shown as the shaded gray region). The x-axis is the initial semi-major axis of DI Herculis, and the y-axis is the effective semi-major axis of a tertiary companion. The region below the green line indicates parameter space for which the Lidov-Kozai timescale is shorter than the 5\,Myr system age (equation \ref{eq:age}). The region below the black line indicates the parameter space for which Lidov-Kozai cycles may occur (equation \ref{eq:migration_condition}), and below the blue line shows where the obliquity may be excited (equation \ref{eq:amax}). The magenta line indicates the dynamical stability criterion, assuming a circular tertiary orbit (equation \ref{eq:stability}); an eccentric orbit would raises the magenta line and reduce the area of the gray region. The horizontal cyan line indicates an RV semi-amplitude of 10 km/s, assuming $\sin I_{\rm p} = 1$. We have assumed a tertiary mass of $1\,\msun$; the allowed parameter space depends only weakly on tertiary mass.}
\label{fig:highemig}
\end{figure}

Figure \ref{fig:highemig} displays the allowed range for the 
perturber parameters of equation~(\ref{eq:perturber_parameter})
as a function of the initial binary semi-major axis.
When these parameters fall in the gray region (and with appropriate inclinations between the orbits) high-eccentricity migration and obliquity growth are possible and consistent
with dynamical stability. For a given value of the initial semi-major axis, the allowed range of perturber parameters spans about an order of magnitude.

As LK-driven high-eccentricity migration proceeds, the inner binary becomes decoupled from the tertiary body. The eccentricity freezes near the maximum value at the point of decoupling, and then the orbit gradually circularizes (see, e.g.,~Fig.~1 of \citealt{fabrycky2007} or Fig.~1 of \citealt{anderson2016}). If DI Herculis is in this latter stage of high-eccentricity migration (in which the eccentricity is no longer varying), the tidal evolution should cause the spin frequency of each star to reach an equilibrium (often called ``pseudo-synchronous'') value \citep[e.g.][]{hut1981}. This is because the stellar spin angular momenta are much smaller than the orbital angular momentum (see Section 3.3 of \citealt{anderson2016} for an analysis of the spin behavior in the regime of small spin angular momentum).

For eccentric and oblique systems, the equilibrium spin is given by equation 4 of \citep{levrard2007}
\be
\frac{\Omega_{\star,\rm eq}}{n} = \frac{1 + \frac{15}{2}e^2 + \frac{45}{8}e^4 + \frac{5}{16}e^6}{\big(1 + 3e^2 + \frac{3}{8} e^4 \big) \big(1 - e^2\big)^{3/2}} \left( \frac{2 \cos \theta}{1 + \cos^2 \theta} \right).
\label{eq:pseudo}
\ee
Inserting the observed values of $e$ and $n$ gives an equilibrium spin period of about 4~days for $\theta = 0^\circ$. For $\theta = 70^\circ$ and $\theta = 80^\circ$, the corresponding spin periods are $\sim 6.6$ and $12$ days. These are much longer than the 1-day observed spin periods (see Fig.~\ref{fig:notertiary}). The observation that the stellar rotation rate is roughly a factor of ten higher than the equilibrium value is evidence against the high-eccentricity-migration scenario for DI Herculis.

Another potential problem is the apparent fine
tuning that would be required for a relatively low-mass tertiary (which is required from the lack of a definitive directly-imaged companion) to drive the high eccentricities needed for migration. In particular, the maximum eccentricity in an LK cycle (which determines whether migration can occur) depends on the initial inclination between the inner and outer orbits. When the angular momentum of the outer binary is much larger than that of the inner binary, large eccentricities in the inner binary can be achieved over a relatively wide range of mutual inclinations. When the angular momentum of the outer binary is on the same order as (or smaller than) that of the inner binary, it is still possible to generate large eccentricities but only over a very narrow range of retrograde mutual inclinations \citep[see Fig.~1 of][]{anderson2017}. Since the total mass of DI Herculis is about $10\,\msun$, it seems unlikely that a tertiary of several solar masses or less could have driven migration without fine-tuning the initial inclination.

Recently, a candidate tertiary companion to DI Herculis was reported by \cite{laos2020}. This report was based on adaptive-optics data in which the image of DI Herculis appeared to deviate slightly from the
expected point-spread function. If real, the tertiary has a separation of about 100\,AU. Examining Fig.~\ref{fig:highemig}, this companion may have driven high-eccentricity migration if the initial semi-major axis of DI Herculis was in the approximate range 3--40\,AU. Further imaging or other observations to confirm or refute the presence of this candidate tertiary star would be helpful.

\section{Conclusion}
\label{sec:conclusion}

We have evaluated the prospects for an inclined tertiary companion to have produced the nearly polar obliquities of the DI Herculis eclipsing binary system. An inclined tertiary may produce spin-orbit misalignment only if the secular torque from the tertiary is sufficiently strong compared to the torques coupling the spins with the inner orbit. Given the current properties of DI Herculis, such a tertiary would need to be close and/or massive (see Fig.~\ref{fig:paramspace}). Sampling over a wide range of tertiary properties and orientations, we show that the obliquity could not have been excited at the present orbital semi-major axis, as the companion would have led to rapid apsidal precession (Fig.~\ref{fig:withtertiary}) and large changes in eccentricity and inclination (Figs.~\ref{fig:timeev} and \ref{fig:delta_e_I}).  We also considered a scenario in which the orbital separation of DI Herculis was initially wider, and subsequently shrank via high-eccentricity tidal migration. Although we find a generous region of parameter space that may allow for both migration and obliquity excitation (see Fig.~\ref{fig:highemig}), we disfavor such a dynamical history due to the lack of pseudo-synchronized stellar spins, as well as the lack of a massive tertiary star (which is likely needed for high-eccentricity migration to be induced without fine tuning). An in-depth quantitative study of high-eccentricity migration in the context of DI Herculis would be useful to evaluate the plausibility of this scenario more definitively.

We have also updated the theoretically-predicted apsidal precession rate (consisting of contributions from general relativity, and tidal and rotational distortion) in the absence of a tertiary companion, using recent photometric detections of the spin period by \cite{liang2022}. This updated apsidal motion rate remains consistent with the observed value obtained by \cite{claret2010} (see Fig.~\ref{fig:notertiary}).

We used the present physical properties of DI Herculis to derive constraints on the properties of a tertiary star that could
have produced spin-orbit misalignment, but in the past, these properties may have been different. For example, the stellar radii were likely larger in the past, as the binary contracted during the pre-main-sequence phase. By itself, a larger radius would increase the quadrupole moment of each star, which would in turn require that the tertiary be even closer to excite the obliquities (and therefore even more easily detected). On the other hand, the stars would have been spinning more slowly. The net effect on the stellar quadrupole moments is therefore uncertain.

We have assumed throughout this work that the binary formed with spin-orbit alignment and that an external torque is required to misalign the stars.  Initial spin-orbit alignment is a natural expectation of star formation for binaries with small orbital separations. However, for initially wider binaries that eventually undergo high-eccentricity migration, the likelihood of primordial misalignment may increase. The influential study of \cite{hale1994} concluded that binaries with orbital separations less than $\sim$\,30\,AU generally exhibit spin-orbit alignment, while those beyond $30$ AU tend to be randomly oriented. However, a recent re-analysis by \cite{justesen2020} found that the available data are insufficient to make inferences about spin-orbit alignment trends. 

For DI Herculis, the calculations we performed (along with the lack of a definite detection of a tertiary star thus far) suggest that we should seek alternative explanations for obliquity excitation or primordial misalignment. \cite{anderson2021} considered a mechanism for exciting obliquities in stellar binaries due to an inclined circumbinary disk. The inclined disk introduces nodal precession of the binary orbit (analogous to a tertiary companion), while each oblate star precesses due to the opposite star. As the disk loses mass through both winds and accretion onto the stars, a secular spin-orbit resonance may be encountered, and the obliquities may be driven towards $90^\circ$ in some circumstances. Such a mechanism (or another mechanism involving a disk and and inward migration) appears to be a more promising explanation than a tertiary companion for the large stellar tilts in DI Herculis.

\acknowledgements
We are grateful to Yan Liang, Roman Rafikov, and Eric Jensen for useful and stimulating discussions. We also thank Gongjie Li for assistance with the GRIT N-body package, and the anonymous referee for comments and feedback. KRA is supported by a Lyman Spitzer, Jr.~Postdoctoral Fellowship at Princeton University.

\bibliography{refs}

\end{document}